\documentclass[aps,prl,twocolumn,floatfix,bibnotes]{revtex4-2}
\usepackage{graphicx}
\usepackage{amsmath}
\usepackage{amssymb}
\usepackage{bm}
\usepackage{color}
\usepackage{dcolumn}
\usepackage{scalerel}
\usepackage{multirow}
\usepackage[caption=false]{subfig}
\usepackage{hyperref}
\usepackage{float}

\begin{document}

\title{Trap-induced ac Zeeman shift of the thorium-229 nuclear clock frequency}

\newcommand{\NIST}{
National Institute of Standards and Technology, Boulder, Colorado 80305, USA
}

\author{K. Beloy}
\email{kyle.beloy@nist.gov}
\affiliation{\NIST}

\date{\today}

\newcommand{\later}{\ensuremath{\spadesuit}}
\newcommand{\pprime}{\ensuremath{{\prime\prime}}}

\begin{abstract}
We examine the effect of a parasitic rf magnetic field, attributed to ion trapping, on the highly anticipated nuclear clock based on $^{229}$Th$^{3+}$ [C. J. Campbell {\it et al.}, Phys.\ Rev.\ Lett.\ 108, 120802 (2012)]. The rf magnetic field induces an ac Zeeman shift to the clock frequency. As we demonstrate, this shift threatens to be the dominant systematic frequency shift for the clock, exceeding other systematic frequency shifts and the projected systematic uncertainty of the clock by orders of magnitude. We propose practical means to suppress or eliminate this shift.
\end{abstract}

\maketitle

\newcommand{\insertfigmanifolds}{
\begin{figure}[t]
\includegraphics[width=246pt]{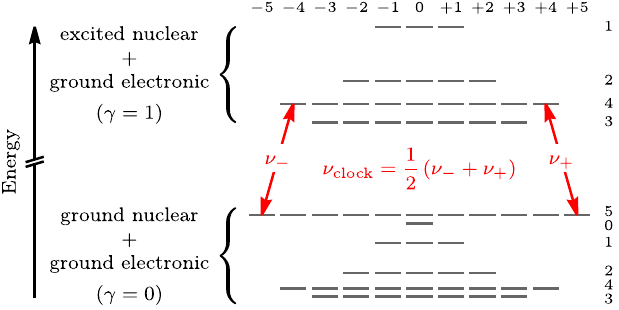}
\caption{Hyperfine manifolds of the $^{229}$Th$^{3+}$ system, with states labeled by $F$ (right) and $m_F$ (top). The clock frequency is defined in terms of the two transitions indicated by arrows. For scale, the curly brackets have a vertical extent of $h\times2.3~\text{GHz}$, where $h=2\pi\hbar$ is Planck's constant.}
\label{Fig:manifolds}
\end{figure}
}

\newcommand{\insertfigbigplot}{
\begin{figure*}[t]
\includegraphics[width=510pt]{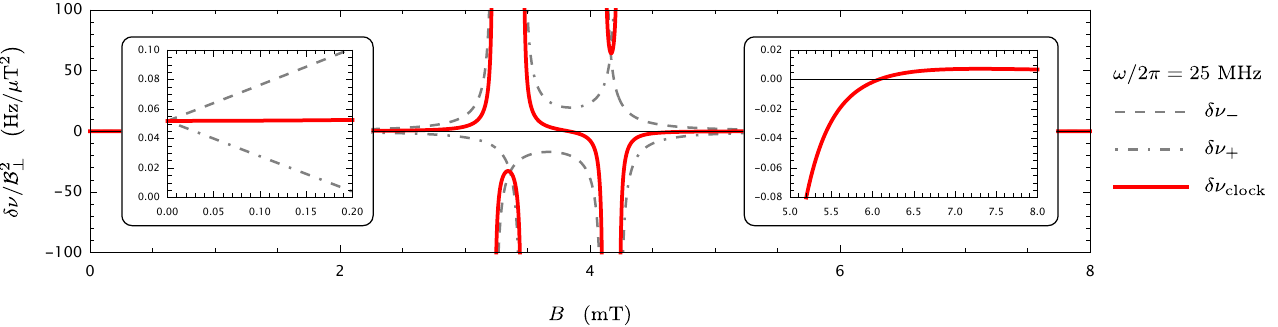}
\caption{Trap-induced ac Zeeman shift to the clock transition frequencies ($\delta\nu\rightarrow\delta\nu_\pm$) and the clock frequency ($\delta\nu\rightarrow\delta\nu_\mathrm{clock}$). The shifts are plotted versus $B$, assuming a trap drive frequency $\omega/2\pi=25~\text{MHz}$. The main plot displays resonant behavior attributed to the $\Delta F=0$ intermediate states. The inset on the left highlights the low-$B$ regime, where $\delta\nu_\mathrm{clock}$ is effectively constant. The inset on the right highlights a zero crossing for $\delta\nu_\mathrm{clock}$ at $B\approx6.1$~mT. }
\label{Fig:bigplot}
\end{figure*}
}

\newcommand{\inserttabparams}{
\begin{table}[b]
\caption{Hyperfine manifold parameters. $A_\mathrm{hfs}$ and $B_\mathrm{hfs}$ are in units of $h\times\text{MHz}$; the other quantities are dimensionless. Refer to the text for references.}
\label{Tab:data}
\begin{ruledtabular}
\begin{tabular}{ccccccc}
$\gamma$ & $I$ & $J$ & $A_\mathrm{hfs}$ & $B_\mathrm{hfs}$ & $g_I$ & $g_J$
\\
\hline
\vspace{-3mm}\\
$0$	& $5/2$	& $5/2$		& $82.2$	& $2269$	& $0.147$		& $6/7$	\\
$1$	& $3/2$	& $5/2$		& $-142$	& $1259$	& $-0.255$		& $6/7$
\end{tabular}
\end{ruledtabular}
\end{table}
}

\newcommand{\insertfigbetaB}{
\begin{figure*}[t]
\includegraphics[width=510pt]{beta0.pdf}%
\caption{$\beta_{\pm1}^\mathrm{x}(\omega)$ for the $^{229}$Th$^{3+}$ clock ($\mathrm{x}=\mathrm{s,t,c}$). Left and center panels: $\beta_{-1}^\mathrm{s}(\omega)$ for the $\gamma=0$ manifold (dashed curves) and the $\gamma=1$ manifold (dotted curves) and the corresponding $\beta_{-1}^\mathrm{t}(\omega)$ (full curves). Right panel: $\beta_{-1}^\mathrm{t}(\omega)$ (curves with respective color from the left and center panels) and $\beta_{-1}^\mathrm{c}(\omega)$ (yellow curve). While the plotted curves are for $\beta_{-1}^\mathrm{x}(\omega)$, curves for $\beta_{+1}^\mathrm{x}(\omega)$ can be inferred from these by recalling that $\beta_q^\mathrm{s}(\omega)$ is invariant under the simultaneous substitutions $m_F\rightarrow-m_F$ and $q\rightarrow-q$.}
\label{Fig:betaB0}
\end{figure*}
}

\newcommand{\insertfigbetaBB}{
\begin{figure*}[t]
\includegraphics[width=510pt]{beta50.pdf}%
\caption{$\beta_{\pm1}^\mathrm{x}(\omega)$ for the $^{229}$Th$^{3+}$ clock ($\mathrm{x}=\mathrm{s,t,c}$), where effects of the dc magnetic field on the states and energies are taken into account. $B=50$~mT is assumed for these curves. Left and center panels: $\beta_{\pm1}^\mathrm{s}(\omega)$ for the $\gamma=0$ manifold (dashed curves) and the $\gamma=1$ manifold (dotted curves) and the corresponding $\beta_{\pm1}^\mathrm{t}(\omega)$ (full curves). Right panels: $\beta_{\pm1}^\mathrm{t}(\omega)$ (curves with respective color from the left and center panels) and $\beta_{\pm1}^\mathrm{c}(\omega)$ (yellow curve). Note the reduced vertical scale for the right panels.}
\label{Fig:betaB50}
\end{figure*}
}

\newcommand{\inserttabintstates}{
\begin{table}[b]
\caption{Clock states and the two intermediate states that contribute to $\beta_\perp$ for each clock state (respective columns). The pairs of numbers specify $F,m_F$.}
\label{Tab:intstate}
\begin{ruledtabular}
\begin{tabular}{lcccc}
& \multicolumn{2}{c}{$\gamma=0$}
& \multicolumn{2}{c}{$\gamma=1$} \\
\cline{2-3}\cline{4-5}
\vspace{-3mm}\\
clock state
& $5,-5$		& $5,+5$		& $4,-4$		& $4,+4$		\\
$\Delta F=0$ int.~state
& $5,-4$		& $5,+4$		& $4,-3$		& $4,+3$		\\
$\Delta F=-1$ int.~state
& $4,-4$		& $4,+4$		& $3,-3$		& $3,+3$		\\
\end{tabular}
\end{ruledtabular}
\end{table}
}

Insensitivity to external influences is a desirable attribute for any frequency reference or timekeeping device. A high-quality wristwatch, for instance, is one that boasts insensitivity to temperature variations caused by body heat or the ambient environment. Atomic clocks exploit the intrinsic insensitivity of atoms to external influences. Still, among the multitude of atomic systems available, some are more favorable than others. With shrewd choices of atomic systems, modern-day atomic clocks have achieved total systematic uncertainties at the low-$10^{-18}$ level~\cite{HunSanLip16,McGZhaFas18,
BreCheHan19,BotKedOel19,HuaZhaZen22,LudBoyYe15} (throughout, shifts and uncertainties given without units are understood to be in units of the respective clock frequency). Continued efforts on these clocks will presumably be rewarded with even lower systematic uncertainties. Alternatively, other atomic systems could offer unprecedented levels of clock accuracy~\cite{CamRadKuz12,KozSafCre18,Bel21,LeiPorChe22arXiv}. One intriguing candidate is $^{229}$Th$^{3+}$. The thorium-229 nucleus possesses a metastable excited state with an anomalously low excitation energy, opening up the possibility of a clock based on a nuclear excitation~\cite{PeiTam03}. For practical realization, the single-valence system $^{229}$Th$^{3+}$ offers advantages over, e.g., the bare nuclear system $^{229}$Th$^{90+}$. Campbell {\it et al.}\ outlined a promising architecture for a $^{229}$Th$^{3+}$ clock~\cite{CamRadKuz12}. With their approach, the clock frequency is largely insensitive to interactions between the atomic electrons and external electromagnetic fields. The benign presence of the electrons, in this respect, justifies the term ``nuclear clock'' often used colloquially and within the literature. By examining important systematic effects, Campbell {\it et al.}\ arrived at a projected systematic uncertainty of $1.5\times10^{-19}$ for the clock. The promise of unprecedented clock accuracy, coupled with the leverage such accuracy affords tests of fundamental physics~\cite{Fla06,FadBerFla20}, has stimulated intense research into the development of this clock~\cite{VonSei20,PeiSchSaf21}.

In this Letter, we identify a critical systematic frequency shift for the prospective $^{229}$Th$^{3+}$ clock. For ion clocks, ion confinement is typically realized with an rf electric field~\cite{Pau90}. The rf electric field, however, is accompanied by an rf magnetic field, with typical field strengths of order $\mu$T at the ion. This parasitic rf magnetic field induces second order ac Zeeman shifts to the atomic energy levels and, consequently, the clock frequency~\cite{BerMilBer98,GanMasTse18}. Here we examine this effect in the $^{229}$Th$^{3+}$ clock. In contrast to other systematic frequency shifts, the electrons' presence is hardly benign, with the $^{229}$Th$^{3+}$ clock holding no advantage over more-traditional ion clocks based on electronic excitations. We find the clock's sensitivity to rf magnetic fields to be comparable to or larger than optical ion clocks previously considered~\cite{GanMasTse18}, with the corresponding shift likely to be the dominant systematic frequency shift for the clock. We further propose means to suppress or eliminate this shift, restoring the clock's insensitivity to external perturbations.

An atomic system is a composition of nucleons and electrons. The energy eigenstates can be described in terms of the state of the nucleus, the state of the electrons surrounding the nucleus, and the total (nuclear + electronic) angular momentum of the system. The coarse energy spectrum is determined by the nuclear state + electronic state, which we encapsulate with the quantum number $\gamma$. Each $\gamma$ identifies a hyperfine manifold. The total angular momentum is specified by the conventional quantum numbers $F$ and $m_F$. These quantum numbers distinguish the states within each manifold, with the energies depending on $F$ but not $m_F$. In atomic clocks and other applications, a dc magnetic field $\mathbf{B}$ defines the quantization axis and lifts the degeneracy in $m_F$.

Figure~\ref{Fig:manifolds} illustrates two manifolds of the $^{229}$Th$^{3+}$ system, to which we assign the values $\gamma=0,1$. The $\gamma=0$ manifold is associated with the ground nuclear state + ground electronic state, while the $\gamma=1$ manifold is associated with the excited nuclear state + ground electronic state. Here ``excited nuclear state'' refers to the low-lying metastable state of the nucleus with an energy of $\approx\!8$~eV~\cite{BecBecBei07,SeiVonBil19,
YamMurHay19,SikGeiHen20,mynote}. In conventional notation, the ground electronic state is $[\mathrm{Rn}]5f~{^2F_{5/2}}$. The clock proposal of Ref.~\cite{CamRadKuz12} invokes the stretched states of these manifolds. For a given manifold, ``stretched states'' refers to the pair of states with maximum $|m_F|$. We refer to the stretched states of the $\gamma=0,1$ manifolds as the clock states. Two transitions connecting these clock states are indicated in Fig.~\ref{Fig:manifolds}, which we refer to as the clock transitions. The transition frequencies of the clock transitions are denoted as $\nu_-$ and $\nu_+$. The proposed clock operation involves alternately driving the two clock transitions, with the clock frequency taken as $\nu_\mathrm{clock}=\left(\nu_-+\nu_+\right)/2$. Following this prescription, $\nu_\mathrm{clock}$ is largely insensitive to interactions between the electrons and external electromagnetic fields (meanwhile, interactions between the nucleons and external electromagnetic fields are naturally suppressed). This insensitivity encompasses ac Stark shifts, electric quadrupole shifts, and dc Zeeman shifts~\cite{CamRadKuz12}. Other transitions, or combinations of transitions, between the $\gamma=0,1$ manifolds generally do not offer such insensitivity.

\insertfigmanifolds%

We now consider the atomic response to an ac magnetic field $\bm{\mathcal{B}}\cos\left(\omega t\right)$, where $\bm{\mathcal{B}}$ and $\omega$ are the vector amplitude and angular frequency of the field, respectively, and $t$ denotes time. Linear polarization is assumed in accordance with Ref.~\cite{GanMasTse18} (see additional discussion below). The ac magnetic field shifts the atomic energy levels according to $\delta E=-\frac{1}{4}\left(\mathcal{B}_{||}^2\beta_{||}+\mathcal{B}_\perp^2\beta_\perp\right)$, where $\bm{\mathcal{B}}$ is partitioned into parallel and perpendicular components relative to the dc magnetic field, $\bm{\mathcal{B}}=\bm{\mathcal{B}}_{||}+\bm{\mathcal{B}}_\perp$, with respective magnitudes $\mathcal{B}_{||}\equiv\left|\bm{\mathcal{B}}_{||}\right|$ and $\mathcal{B}_\perp\equiv\left|\bm{\mathcal{B}}_\perp\right|$. From second order perturbation theory, the corresponding magnetic dipole polarizabilities $\beta_{||}$ and $\beta_\perp$ satisfy
\begin{gather*}
\beta_{||}=-\sum_{\gamma^\prime F^\prime m_F^\prime}\sum_{\sigma}\frac{\left|\langle\Psi_{\gamma Fm_F}|\mu_0|\Psi_{\gamma^\prime F^\prime m_F^\prime}\rangle\right|^2}{E_{\gamma Fm_F}-E_{\gamma^\prime F^\prime m_F^\prime}+\sigma\hbar\omega},
\\
\beta_\perp=-\frac{1}{2}\sum_{\gamma^\prime F^\prime m_F^\prime}\sum_{\sigma q}\frac{\left|\langle\Psi_{\gamma Fm_F}|\mu_q|\Psi_{\gamma^\prime F^\prime m_F^\prime}\rangle\right|^2}{E_{\gamma Fm_F}-E_{\gamma^\prime F^\prime m_F^\prime}+\sigma\hbar\omega},
\end{gather*}
where $\sigma$ and $q$ take on the values $\pm1$ in the summations. Here $\mu_q$ is the $q$-th spherical component of the magnetic dipole operator $\bm{\mu}$, with the $q=0$ axis aligned with $\mathbf{B}$. For our purposes, we regard $|\Psi_{\gamma Fm_F}\rangle$ and $E_{\gamma Fm_F}$ as the states and energies of the atomic system inclusive of the dc Zeeman interaction (i.e., interaction with $\mathbf{B}$), with the unprimed set of quantum numbers in the expressions identifying with the state under consideration. The quantum numbers are understood to adiabatically connect to those of the unperturbed atomic system in the limit $B\rightarrow0$, where $B\equiv\left|\mathbf{B}\right|$. In our treatment, the polarizabilities $\beta_{||}$ and $\beta_\perp$ depend on both $B$ and $\omega$.

Intermediate-state contributions to the polarizabilities can be categorized as either intramanifold ($\gamma^\prime=\gamma$) or extramanifold ($\gamma^\prime\neq\gamma$). For the $B$ and $\omega$ of interest here, the extramanifold contributions are small and largely cancel between states of the respective clock transitions. Meanwhile, the intramanifold contributions are generally much larger, and no appreciable cancellation should be expected. In the remainder, we restrict our consideration to the intramanifold contributions. Regarding the extramanifold contributions, we note that a similar cancellation occurs for the ac Stark effect. Important systematic frequency shifts are correspondingly suppressed, such as the blackbody radiation shift~\cite{CamRadKuz12}. A critical contrast should be appreciated here. For the ac Stark effect, intramanifold contributions are absent due to parity selection rules. For the ac Zeeman effect, on the other hand, the intramanifold contributions dominate.

The dc Zeeman interaction preserves axial symmetry of the system. As such, the matrix element $\langle\Psi_{\gamma Fm_F}|\mu_q|\Psi_{\gamma^\prime F^\prime m_F^\prime}\rangle$ is subject to the selection rule $\Delta m_F=-q$, where $\Delta m_F\equiv m_F^\prime-m_F$. This restricts the intermediate states that contribute to the polarizabilities. For the clock states, this implies $\beta_{||}=0$, since stretched states have distinct $m_F$ quantum numbers within their manifold. Meanwhile, for each clock state, only two intermediate states contribute to $\beta_\perp$. These two states can be distinguished by $\Delta F=0,-1$, where $\Delta F\equiv F^\prime-F$. We recall that the dc Zeeman interaction lifts the degeneracy between the $\Delta F=0$ intermediate state and its respective clock state. Table~\ref{Tab:intstate} specifies the intermediate states associated with each clock state.

\inserttabintstates%

We conclude from the preceding discussion that $\delta\nu\propto\mathcal{B}_\perp^2$, where $\delta\nu$ represents the ac Zeeman shift to one of the clock transition frequencies ($\delta\nu\rightarrow\delta\nu_\pm$) or the clock frequency ($\delta\nu\rightarrow\delta\nu_\mathrm{clock}$). The proportionality factor is $-1/4h$ times the appropriate combination of clock state polarizabilities $\beta_\perp$. In ion clocks, the parasitic rf magnetic field is not  carefully controlled~\cite{BerMilBer98,GanMasTse18,BreCheBel19}. Thus, the experimental condition $\mathcal{B}_\perp=0$ cannot simply be assumed, as was effectively done in Ref.~\cite{CamRadKuz12}. We further note that $\mathcal{B}_\perp$ is invariant with respect to inversion of $\mathbf{B}$. Moreover, an average over three mutually orthogonal $\mathbf{B}$ directions amounts to replacing $\mathcal{B}_\perp^2$ with the scalar quantity $(2/3)\mathcal{B}^2$, where $\mathcal{B}\equiv\left|\bm{\mathcal{B}}\right|$. Thus, while capable of eliminating certain shifts in ion clocks~\cite{Ita00}, such averaging does not eliminate the ac Zeeman shift~\cite{GanMasTse18}.

Evaluating the polarizabilities requires intramanifold energy differences and matrix elements. To this end, we introduce a model space for each manifold given by the product states $|Im_I\rangle|Jm_J\rangle$. Here $I$ and $m_I$ ($J$ and $m_J$) are conventional quantum numbers specifying the nuclear (electronic) angular momentum. $I$ and $J$ are manifold-specific, while all compatible values of $m_I$ and $m_J$ define the $\left[(2I+1)(2J+1)\right]$-dimensional model space. Within the model space, the hyperfine interaction has the form
\begin{gather*}
H_\mathrm{hfi}=
A_\mathrm{hfs}\left(\bm{\mathcal{I}}\cdot\bm{\mathcal{J}}\right)
+B_\mathrm{hfs}
\left(
\left\{\bm{\mathcal{I}}\otimes\bm{\mathcal{I}}\right\}_2
\cdot
\left\{\bm{\mathcal{J}}\otimes\bm{\mathcal{J}}\right\}_2
\right)/X,
\end{gather*}
where $A_\mathrm{hfs}$ and $B_\mathrm{hfs}$ are hyperfine constants (taken here to have dimension of energy), $X=(2/3)I(2I-1)J(2J-1)$, and $\bm{\mathcal{I}}$ ($\bm{\mathcal{J}}$) is the nuclear (electronic) angular momentum operator divided by $\hbar$. Here $\left\{\bm{\mathcal{I}}\otimes\bm{\mathcal{I}}\right\}_2$ is the rank-2 tensor operator formed by coupling $\bm{\mathcal{I}}$ with itself (likewise for $\bm{\mathcal{J}}$), while dots signify a scalar product~\cite{VarMosKhe88}. The eigenstates of $H_\mathrm{hfi}$ are $|(IJ)Fm_F\rangle=\sum_{m_Im_J}C^{Fm_F}_{Im_IJm_J}|Im_I\rangle|Jm_J\rangle$, where $C^{Fm_F}_{Im_IJm_J}$ is a Clebsch-Gordan coefficient. Here $F$ and $m_F$ are the same quantum numbers as above, characterizing the total angular momentum and distinguishing states of the manifold. 
Within the model space, $\bm{\mu}$ has the form
\begin{gather}
\bm{\mu}=g_I\mu_N\bm{\mathcal{I}}-g_J\mu_B\bm{\mathcal{J}},
\label{Eq:MSmu}
\end{gather}
where $g_I$ ($g_J$) is the nuclear (electronic) $g$-factor and $\mu_N$ ($\mu_B$) is the nuclear (Bohr) magneton. The Hamiltonian $H_\mathrm{tot}\equiv H_\mathrm{hfi}-\bm{\mu}\cdot\mathbf{B}$ incorporates the dc Zeeman interaction. In our approach, the intramanifold energy differences are obtained from the eigenvalues of $H_\mathrm{tot}$, while the intramanifold matrix elements are obtained by evaluating $\bm{\mu}$, given by Eq.~(\ref{Eq:MSmu}), between the eigenstates of $H_\mathrm{tot}$. Note that this approach treats the dc Zeeman interaction nonperturbatively within the model space, on equal footing with the hyperfine interaction~\cite{Der10a,Der10b,ChiNelOlm11,BakSch14,ArnBar16}.

\inserttabparams%

Table~\ref{Tab:data} compiles the manifold parameters for the $\gamma=0,1$ manifolds. For the $\gamma=0$ manifold, $A_\mathrm{hfs}$ and $B_\mathrm{hfs}$ are experimental results from Ref.~\cite{CamRadKuz11}, while $g_I$ is from $A_\mathrm{hfs}$ combined with electronic structure calculations~\cite{SafSafRad13,PorSafKoz21,LiQiaTan21}. For the $\gamma=1$ manifold, $A_\mathrm{hfs}$, $B_\mathrm{hfs}$, and $g_I$ are obtained from the respective $\gamma=0$ parameters, together with experimental nuclear-moment ratios given in Ref.~\cite{ThiOkhGlo18}. As both manifolds are associated with the same electronic state, $g_J$ is a common parameter. We use the nonrelativistic value for a ${^2}F_{5/2}$ electronic state, $g_J=6/7$. An {\it ab initio} relativistic many-body calculation suggests that relativistic corrections to $g_J$ are sub-percent. Uncertainties are not included in Table~\ref{Tab:data}, as our essential conclusions are insensitive to the exact values of these parameters.

The clock proposal of Ref.~\cite{CamRadKuz12} assumes a weak dc magnetic field and a trap drive frequency of 25~MHz. With a reasonable interpretation of ``weak,'' this corresponds to the operational regime $\mu_BB\ll\hbar\omega\ll\Delta E_\mathrm{hfs}$, where $\Delta E_\mathrm{hfs}$ is the hyperfine energy splitting between the $\Delta F=-1$ intermediate state and its respective clock state. Namely, $\Delta E_\mathrm{hfs}/h=(1770,438)$~MHz for the $\gamma=0,1$ manifolds, respectively. Within this operational regime, the $\Delta F=0$ intermediate-state contributions are suppressed, and $\delta\nu_\mathrm{clock}$ is only weakly dependent on $B$ and $\omega$. For instance, using the above prescription, we find that $\delta\nu_\mathrm{clock}/\mathcal{B}_\perp^2$ varies by only 2\% over the parameter space $0\leq B\leq10~\mu\text{T}$ and $1~\text{MHz}\leq\omega/2\pi\leq50~\text{MHz}$. Using $\mathcal{B}_\perp=1~\mu\text{T}$ as a point of reference, the clock shift evaluates to $2.6\times10^{-17}$. This exceeds all other anticipated systematic frequency shifts and the projected systematic uncertainty of the clock by more than two orders of magnitude~\cite{CamRadKuz12}. Given the quadratic dependence on $\mathcal{B}_\perp$, the shift could be appreciably larger in practice. For example, using $\mathcal{B}_\perp=4.5~\mu\text{T}$ as measured in Ref.~\cite{GanMasTse18}, the shift evaluates to $5.3\times10^{-16}$. In any case, it is clear that the performance of the $^{229}$Th$^{3+}$ clock will depend critically on the management of this shift.

The Barrett group in Singapore has developed a spectroscopic technique for evaluating $\mathcal{B}_\perp$, which involves measuring Autler-Townes line splittings induced by the parasitic rf magnetic field~\cite{GanMasTse18}. Using this technique, they have demonstrated evaluation of $\mathcal{B}_\perp$ at the part-per-thousand (ppt) level, with observed values of $\mathcal{B}_\perp\approx1~\mu\text{T}$~\cite{GanMasTse18,ZhiArnKae20,ArnKaeCha20,KaeTanZha20}. A moderate dc magnetic field is required, $B\sim\hbar\omega/\mu_B\sim\text{mT}$, such to tune a dc Zeeman splitting into resonance with the trap drive frequency. This technique could be applied to $^{229}$Th$^{3+}$, though precise extraction of $\mathcal{B}_\perp$ would be somewhat complicated by prominent intramanifold state-mixing induced by the dc Zeeman interaction. To enable sympathetic cooling, Ref.~\cite{CamRadKuz12} suggests trapping a $^{232}$Th$^{3+}$ ion together with the $^{229}$Th$^{3+}$ ion, in which case the (hyperfine-free) $^{232}$Th$^{3+}$ ion may alternatively be used for $\mathcal{B}_\perp$ diagnostics.


\insertfigbigplot

The Singapore technique is a viable option for addressing the trap-induced ac Zeeman shift. 
While Ref.~\cite{ArnKaeCha20} has demonstrated sub-part-per-thousand instability of $\mathcal{B}_\perp$ on the one-hour timescale, periodic reassessments of $\mathcal{B}_\perp$ may be necessary on longer timescales to accommodate, e.g., drifts in the dc magnetic field direction. Given fundamental quantum projection noise associated with a single ion~\cite{ItaBerBol93}, we note that multiple days worth of averaging will likely be required to procure statistical uncertainty at or below $1\times10^{-19}$. Moreover, instability of $\mathcal{B}_\perp$ could be exacerbated for operation outside of a laboratory setting, as could be beneficial for tests of fundamental physics~\cite{SafBudDeM18}. In any case, sensitivity to the parasitic rf magnetic field represents a deviation from the original premise of the ``nuclear'' clock, which is exceptional insensitivity to environmental perturbations. Below we present strategies to restore this insensitivity by leveraging the dependence of the shift on $B$ and $\omega$. The Singapore technique could be a useful complement to these strategies, though with significantly relaxed demands on the control or assessment of $\mathcal{B}_\perp$. 


Figure~\ref{Fig:bigplot} plots $\delta\nu/\mathcal{B}_\perp^2$ as a function of $B$, where $\delta\nu\rightarrow\delta\nu_\pm$ or $\delta\nu\rightarrow\delta\nu_\mathrm{clock}$. A trap drive frequency of $\omega/2\pi=25$~MHz is assumed. The main plot captures resonant behavior attributed to the $\Delta F=0$ intermediate states. We observe two zero crossings for $\delta\nu_\mathrm{clock}$, one residing within the resonant structure at $B\approx3.8$~mT and another residing outside of the resonant structure at $B\approx6.1$~mT (see right inset of Fig.~\ref{Fig:bigplot}). By operating at one of these values of $B$, the clock becomes insensitive to the parasitic rf magnetic field (and, likewise, the direction of the dc magnetic field). From a practical standpoint, the outer zero-crossing is more favorable, as the shifts $\delta\nu_\pm$ and the slope of $\delta\nu_\mathrm{clock}$ with respect to $B$ are substantially smaller. We denote the value of $B$ at this zero crossing as $B^*$. The precise value of $B^*$ can be determined by taking differential measurements of $\nu_\mathrm{clock}$ at different $B$, where $\mathcal{B}_\perp$ is varied in the differential measurements (e.g., by changing the trapping power~\cite{BerMilBer98,BreCheBel19}). This would be analogous to experimental determinations of ``magic'' wavelengths in optical lattice clocks~\cite{TakHonHig05}. Alternatively, the manifold parameters could be precisely measured, and $B^*$ could be evaluated using the theory implemented in this work. We emphasize that $B^*$ depends on the trap drive frequency. By exploring trap drive frequencies in the range $1~\text{MHz}\leq\omega/2\pi\leq100~\text{MHz}$, we observe an approximate proportionality $B^*\propto\omega$. We assume $\omega/2\pi=25~$MHz below, except where stated otherwise.

In practice, deviations $\delta B=B-B^*$ would lead to a residual clock shift $\delta\nu_\mathrm{clock}\approx\left(21~\mu\text{Hz}/\mu\text{T}^3\right)\mathcal{B}_\perp^2\delta B$. The frequency difference $\Delta\nu\equiv\nu_+-\nu_-$ could be used to monitor and stabilize $B$. Note that evaluation of $\Delta\nu$ comes at no cost, since clock operation already incorporates the $\nu_\pm$ measurements. The dominant contribution to $\Delta\nu$ stems from the bare nuclear contribution~\cite{CamRadKuz12}, with $\Delta\nu\approx\left(11~\text{kHz}/\text{mT}\right)B$. However, the ac Zeeman shifts $\delta\nu_\pm$ would also contribute to $\Delta\nu$, being manifested as a deviation $\delta B\propto\mathcal{B}_\perp^2$ and a residual clock shift $\delta\nu_\mathrm{clock}\propto\mathcal{B}_\perp^4$. For $\mathcal{B}_\perp=1~\mu$T, the residual shift evaluates to a negligible value of $1\times10^{-21}$. However, the quartic scaling can make it relevant for values of $\mathcal{B}_\perp$ not much larger. Thus, to utilize $\Delta\nu$, it may be necessary to ensure that $\mathcal{B}_\perp$ remains below a threshold value, such as a few $\mu$T. Alternatively, rf spectroscopy could be performed on either $^{229}$Th$^{3+}$ or $^{232}$Th$^{3+}$ to periodically assess $B$, with negligible influence from the ac Zeeman shift.

In accordance with Ref.~\cite{GanMasTse18}, linear polarization has been assumed for the parasitic rf magnetic field. Reference~\cite{GanMasTse18} argues that nonlinear polarization would be accompanied by significant, detrimental micromotion. In subsequent works, this assumption appears to be validated at the ppt level~\cite{ArnKaeCha20,KaeTanZha20}. Assuming $\mathcal{B}_\perp\approx1~\mu$T with an out-of-phase component of $\lesssim1$~ppt, the residual clock shift due to nonlinear polarization is constrained to below $10^{-20}$ when operating at $B^*$.

Finally, we propose an alternative strategy for managing the trap-induced ac Zeeman shift. In the operational regime $\mu_BB\gg\hbar\omega$, the frequency shifts $\delta\nu_\pm$ and $\delta\nu_\mathrm{clock}$ become highly suppressed due to a high degree of cancellation between the clock states. For instance, with $\omega/2\pi=8$~MHz and $B=50$~mT, $\delta\nu_\mathrm{clock}$ is suppressed by a factor of $\approx\!3000$ relative to the low-$B$ case, corresponding to a clock shift of $<\!10^{-20}$ for $\mathcal{B}_\perp=1~\mu$T. A plot encompassing a range of $B$ and $\omega$ values in this regime is provided in the Supplemental Material~\cite{SMcite}. We note that the second order dc Zeeman shift to $\delta\nu_\mathrm{clock}$ is highly suppressed for the $^{229}$Th$^{3+}$ clock~\cite{CamRadKuz12}. Thus, even with a large value of $B$, the second order dc Zeeman shift is not expected to be problematic.

In conclusion, we have demonstrated that the parasitic rf magnetic field associated with ion trapping will be a critical systematic frequency shift for the $^{229}$Th$^{3+}$ clock proposed in Ref.~\cite{CamRadKuz12}. In contrast to other important systematic frequency shifts such as ac Stark shifts (attributed to trapping fields, blackbody radiation, the probe laser, and the cooling laser), the electric quadrupole shift, and first- and second-order dc Zeeman shifts, the role of the electrons is hardly benign, with the resulting sensitivity to the rf magnetic field being comparable to or larger than other optical ion clocks previously considered~\cite{GanMasTse18}. In particular, this shift is likely to exceed all other systematic frequency shifts and the projected systematic uncertainty of the clock by orders of magnitude. We have further proposed strategies to eliminate or suppress the shift, restoring the prospects for a ``nuclear'' clock with exceptional insensitivity to external perturbations.

\begin{acknowledgments}
The author thanks Y.~Hassan and A.~G.~Radnaev for their careful reading of the manuscript. This work was supported by the National Institute of Standards and Technology/Physical Measurement Laboratory, an agency of the U.S.\ government, and is not subject to U.S.\ copyright.
\end{acknowledgments}


%

\end{document}